\documentclass[runningheads]{llncs}
\usepackage{graphicx}
\usepackage{amsmath,amssymb} 
\usepackage{color}
\usepackage{algorithm,algorithmic}
\usepackage[colorlinks,bookmarksopen,bookmarksnumbered,allcolors=gray]{hyperref}

\usepackage[width=122mm,left=12mm,paperwidth=146mm,height=193mm,top=12mm,paperheight=217mm]{geometry}

\usepackage{xcolor}
\definecolor{mygreen}{HTML}{55d400}

\begin{document}
\title{Multi--Scale Recursive and Perception--Distortion Controllable \\ Image Super--Resolution}

\titlerunning{Multi--Scale Recursive and Perception--Distortion Controllable Image SR}
%
\author{
    Pablo Navarrete Michelini \and
    Dan Zhu \and
    Hanwen Liu
}
%
\authorrunning{P. Navarrete Michelini, D. Zhu and H. Liu}
%

\institute{BOE Technology Group, Co., Ltd.\\ \email{\{pnavarre,zhudan,liuhanwen\}@boe.com.cn}}
\maketitle              
\begin{figure}[h]
  \centering
  \includegraphics[width=0.8\linewidth]{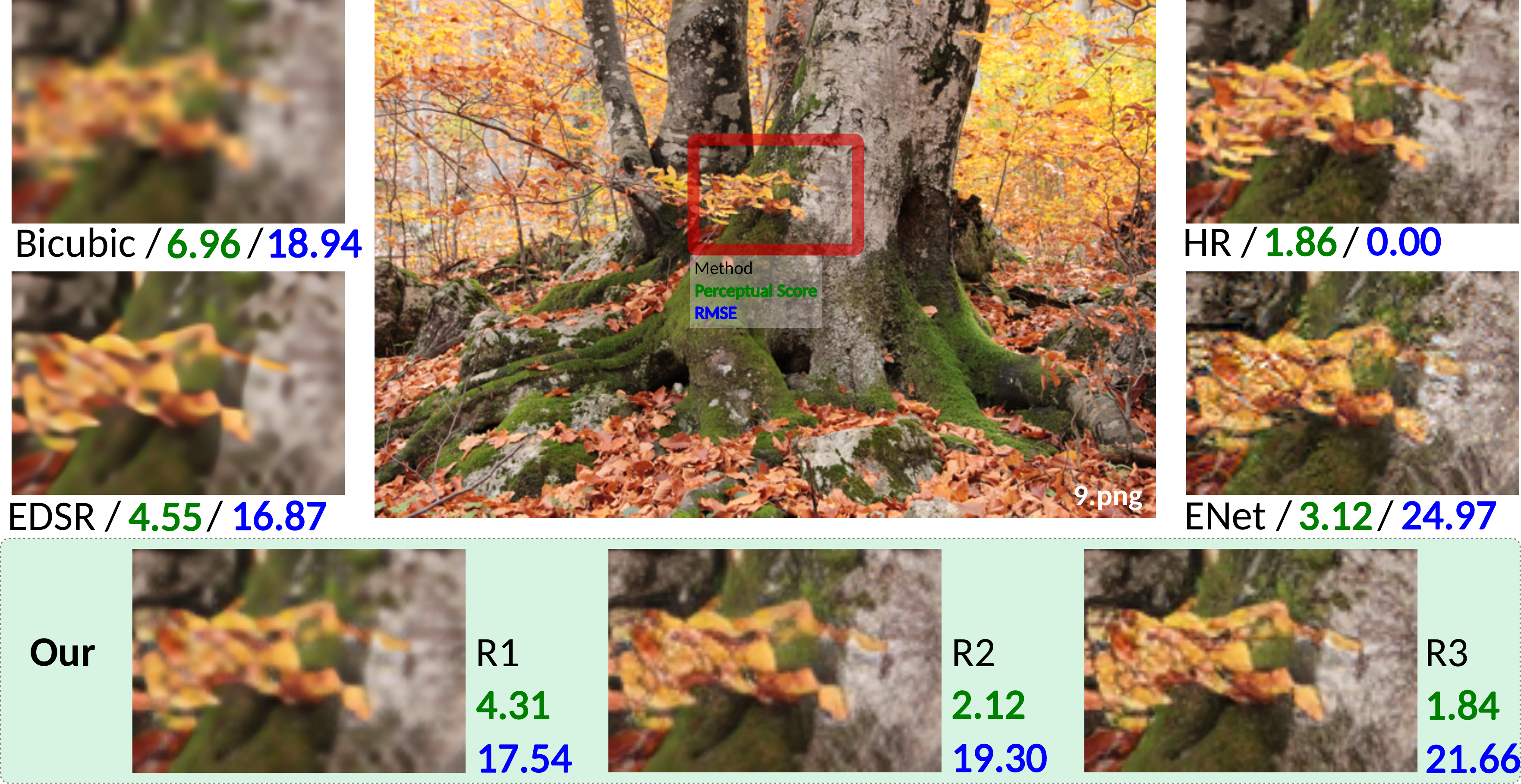}
  \caption{Our G--MGBP super--resolution improves the perceptual quality of low distortion systems like EDSR\cite{Lim_2017_CVPR_Workshops} (with slightly higher RMSE), as well as baseline systems like EnhanceNet\cite{Sajjadi_2017_ICCV} (with significantly lower RMSE). Its perceptual scores are similar to the original images showing its effectiveness for ECCV PIRM--SR Challenge 2018\cite{PIRM-SR}. Code and models are available at \url{https://github.com/pnavarre/pirm-sr-2018}.
}
  \label{fig:teaser}
\end{figure}
\begin{abstract}
We describe our solution for the PIRM Super--Resolution Challenge 2018 where we achieved the \textbf{$\boldsymbol{2^{nd}}$ best perceptual quality} for average $RMSE\leqslant 16$, $5^{th}$ best for $RMSE\leqslant 12.5$, and $7^{th}$ best for $RMSE\leqslant 11.5$. We modify a recently proposed Multi--Grid Back--Projection (MGBP) architecture to work as a generative system with an input parameter that can control the amount of artificial details in the output. We propose a discriminator for adversarial training with the following novel properties: it is multi--scale that resembles a progressive--GAN; it is recursive that balances the architecture of the generator; and it includes a new layer to capture significant statistics of natural images. Finally, we propose a training strategy that avoids conflicts between reconstruction and perceptual losses. Our configuration uses only $281k$ parameters and upscales each image of the competition in $0.2s$ in average.
\end{abstract}
\section{Introduction}
\label{sec:introduction}
We are interested in the problem of single image super--resolution (SR), which is to improve the quality of upscaled images by large factors (e.g. $4\times$) based on examples of pristine high--resolution images. Questions such as the objective meaning of quality, and what characterizes a pristine image, leads us towards different targets. The traditional approach is to focus on the reconstruction of high--resolution images from their downscale versions. We will refer to this target as \emph{distortion} optimization. Alternatively, we can focus on creating upscale images that look as real as natural images to human eyes. We refer to the latter as \emph{perception} optimization. In \cite{Blau_2018_CVPR}, Blau and Michaeli studied the conflicting roles of distortion and perceptual targets for image enhancements problems such as SR. Both targets cannot be achieved at the same time, one must compromise perceptual quality to reduce distortion and vice versa. Here, we are interested in the optimal balance between these two targets.

Our work follows the line of research started by SRCNN\cite{CDong_2014a,CDong_2015a}, which designed SR architectures using convolutional networks. SRCNN focused on a distortion target and it was later improved most notably by EDSR\cite{Lim_2017_CVPR_Workshops} and DBPN\cite{DBPN2018} in NTIRE--SR Challenges\cite{Timofte_2017_CVPR_Workshops,NTIRE2018_SR_Report}. The work on SR architectures with a focus on perceptual targets has been possible thanks to the great progress in Generative Adversarial Networks (GAN)\cite{NIPS2014_5423} and style transfer\cite{Gatys2015c}. It began with SRGAN\cite{DBLP:journals/corr/LedigTHCATTWS16}, which proposed the use of GANs, followed by Johnson\cite{Johnson2016Perceptual}, who proposed a real--time style transfer architecture, and later improved by EnhanceNet\cite{Sajjadi_2017_ICCV}, which combined both approaches. Most recently, the Contextual (CX) loss\cite{mechrez2018contextual} has been used in SR architectures to improve the similarity of feature distributions between artificial and natural images\cite{mechrez2018Learning}. This latest method provides the best benchmark for perceptual quality according to non--reference metrics used in PIRM--SR 2018\cite{PIRM-SR}: Ma\cite{MaYY016} and NIQE\cite{NIQE_2013}.

Our system architecture was inspired by the multi--scale structure of MSLapSR\cite{MSLapSRN}, which we adapted to use Iterative Back--Projections (IBP) in feature space to enforce a downscaling model. In \cite{MGBP} we extended the classic IBP method to multiple scales by using a recursion analogous to the Full Multi--Grid algorithm, which is commonly used as PDE solver\cite{UTrottenberg_2000a}. The system in \cite{MGBP} focused exclusively on a distortion target and now we extend it to perceptual targets.

Our main contributions are:
\begin{itemize}
    \item We propose a novel \textbf{strategy to control the perception--distortion trade-off} in Section \ref{sec:strategy}, which we adopt to design our system.
    \item We introduce \textbf{multi--scale diversity} into our SR architecture design, through random inputs at each upscaling level. These inputs are manipulated by the network in a recursive fashion to generate artificial details at different scales. See Section \ref{sec:generator}.
    \item We propose a novel \textbf{variance--normalization and shift--correlator} (VN+SC) layer that provides meaningful features to the discriminator based upon previous research on the statistics of natural images. See Section \ref{ssec:discr_normalization}.
    \item We propose, to the best of our knowledge, the \textbf{first multi--scale and recursive discriminator} for adversarial training. It is a configuration symmetric to the multi--scale upscaler, therefore it is more effective for adversarial training. It can simultaneously evaluate several upscaling factors, resembling a Progressive GAN\cite{karras2018progressive} in the sense that the optimizer can focus on smaller factors first and then work on larger factors. See Section \ref{ssec:discr_multiscale_recursive}.
    \item We propose a novel \textbf{noise--adaptive training strategy} that can avoid conflicts between reconstruction and perceptual losses, combining loss functions with different random inputs into the system. See Section \ref{sec:training}.
\end{itemize}

\section{Controlling Distortion vs Perceptual Quality}
\label{sec:strategy}
\begin{figure}[t]
    \centering
    \includegraphics[width=0.8\linewidth]{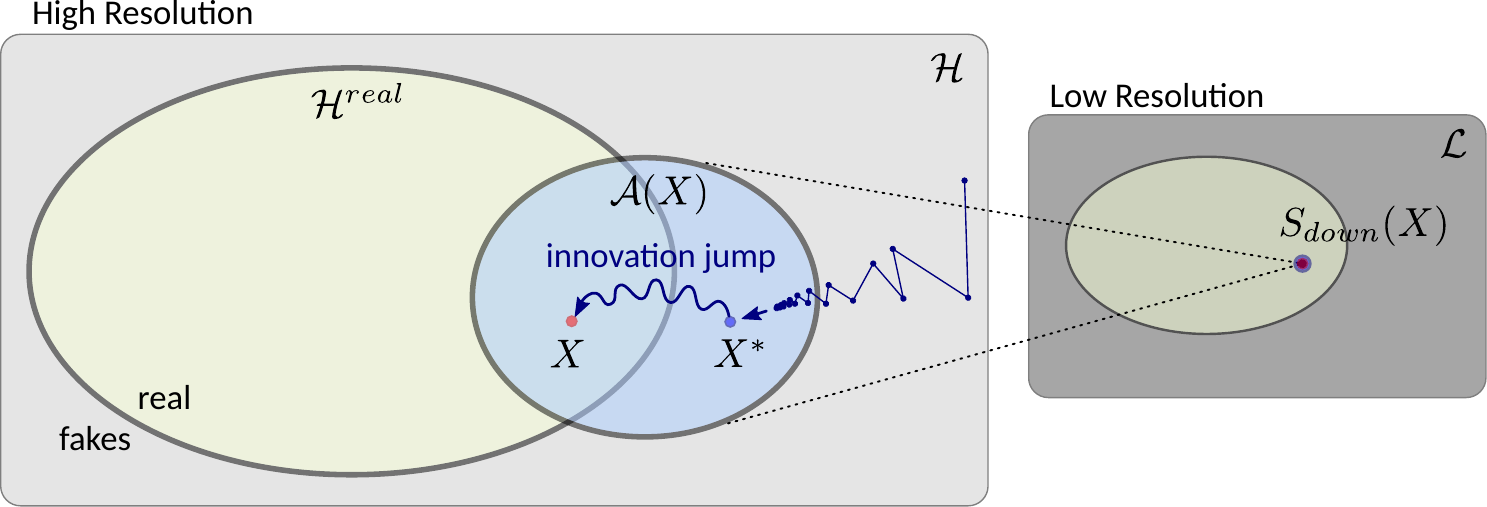}
    \caption{Given a high--resolution image $X$ that looks real, distortion optimization approaches an optimal solution $X^*$ that does not look real because it lacks the random nature of natural images. We can still use $X^*$ as a reference point to move through an \emph{innovation jump} into the set of realistic images. \label{fig:strategy_sets}}
\end{figure}

To better illustrate our target, we present a diagram of image sets in Figure \ref{fig:strategy_sets}. Here, $\mathcal{H}$ is the set of all high--resolution images, $\mathcal{H}^{real}\subset\mathcal{H}$ is the subset of high--resolution images that correspond to natural images, and $\mathcal{L}$ is the set of all low--resolution images. Given an image $X\in\mathcal{H}^{real}$, we are interested in the set of \emph{aliased} images:
\begin{equation}
    \mathcal{A}(X) = \left\{ Y \in \mathcal{H} \quad s.t. \quad S_{down}(Y)=S_{down}(X) \right\} \;,
\end{equation}
where $S_{down}:\mathcal{H}\rightarrow\mathcal{L}$ is a \emph{downscale} operator. We are particularly interested in the set $\mathcal{A}(X)\cap\mathcal{H}^{real}$ of alias images that correspond to real content.

A \emph{distortion} function $\Delta(X,y)$ measures the dissimilarity between a reconstructed image $y$ and the original image $X$. Popular and basic distortion metrics such as L1, L2, PSNR, etc., are sensitive to changes (any minor difference in pixel values would increase the amount of distortion) and are known to have low correlation with human perception\cite{KSeshadrinathan_2010a}. Several distortion metrics have been proposed to approach perceptual quality by emphasizing some differences more than others, either through normalization, feature extraction or other approaches. These include metrics like SSIM\cite{Wang04imagequality}, VIF\cite{VIF} and the VGG content loss\cite{DBLP:journals/corr/JohnsonAL16}. By doing so, correlation with human perception improves according to \cite{KSeshadrinathan_2010a}, but experiments in \cite{Blau_2018_CVPR} show that these metrics still focus more on distortion. More recently, the contextual loss has been proposed to focus more on perceptual quality while maintaining a reasonable level of distortion\cite{mechrez2018contextual}.

The optimal solution of distortion optimization is obtained by:
\begin{equation}
    X^*=\text{argmin}_y \mathbb{E}\left[\Delta(X,y)\right] \;. \label{eq:optimal_distortion}
\end{equation}
The original image $X$ is fixed, and the expected value in \eqref{eq:optimal_distortion} removes any visible randomness in the search variable $y$. But, according to research on the statistics of natural images, randomness plays an essential role in what makes images look real\cite{Ruderman1994}. This is well known for non--reference image quality metrics such as NIQE\cite{NIQE_2013} or BRISQUE\cite{BRISQUE_2012}, and led to a definition of perceptual quality as a distance between probability distributions in \cite{Blau_2018_CVPR}. It is also known that distortion optimization solutions tend to look unreal, as seen in state--of--the--art results from NTIRE--SR Challenges\cite{Timofte_2017_CVPR_Workshops,NTIRE2018_SR_Report}. Common distortion metrics in these challenges (L1 and L2) make the image $X^*$ lose all randomness. We argue that this removal of randomness in $X^*$ is what moves it out of set $\mathcal{H}^{real}$, as we show in Figure \ref{fig:strategy_sets}.

We know that $X\neq X^*$ because $X\in\mathcal{H}^{real}$ and $X^*\notin\mathcal{H}^{real}$ according to our previous discussion. However, distortion optimization can still be useful to generate realistic images. By approaching $X^*$ we are getting closer to $X$. As shown in Figure \ref{fig:strategy_sets}, both $X$ and $X^*$ can be in $\mathcal{A}(X)$. Using a signal processing terminology, the \emph{innovation}\cite{Mitter_1982} is the difference between $X$ and the optimal forecast of that image based on prior information, $X^*$. Most SR architectures take the randomness for the innovation process from the low--resolution input image, which is a valid approach but loses the ability to expose and control it.

In our proposed architecture we add randomness explicitly as noise inputs, so that we can control the amount of innovation in the output. Independent and identically distributed noise will enter the network architecture at different scales, so that each of them can target artificial details of different sizes. Generally speaking, our training strategy will be to approach $X^*$ with zero input noise and any image in $\mathcal{A}(X)\cap\mathcal{H}^{real}$ with unit input noise. By using noise to target perceptual quality, and remove it for the distortion target, we teach the network to \emph{jump} from $X^*$ into $\mathcal{H}^{real}$. With probability one the network cannot hit $X$, but the perceptual target is any image in $\mathcal{A}(X)\cap\mathcal{H}^{real}$.

\begin{algorithm}
    \caption{Generative Multi--Grid Back--Projection (G--MGBP)} \label{alg:gmgbp}
    \begin{tabular}{ll}
        $\boldsymbol{G-MGBP}(X,W,\mu,L)$: & $\boldsymbol{BP^{\mu}_{k}}(u, Y_1,\ldots,Y_{k-1}, noise_1,\ldots,noise_{k-1})$: \\

        \resizebox{.45\textwidth}{!}{
        \begin{minipage}{.5\textwidth}
            \begin{algorithmic}[1]
                \REQUIRE Input image $X$.
                \REQUIRE Noise amplitude $W$.
                \REQUIRE Numbers $\mu$ and $L$.
                \ENSURE $Y_k$, $k = 2,\ldots,L$.

                \STATE $Y_1 = X$
                \STATE $noise_1 = W\cdot\mathcal{N}(0,1)$
                \FOR{$k = 2,\ldots,L$}
                    \STATE $Y_k = \text{ClassicUpscale}(Y_{k-1})$
                    \STATE $d = \text{Downscale}(\text{Analysis}(Y_k))$
                    \STATE $u = \text{Upscale}(\left[Y_{k-1},\; d,\; noise_{k-1}\right])$
                    \STATE $u = BP^{\mu}_{k}(u, Y_1,\ldots,Y_{k-1},$ \\
                           $\quad\quad\quad\quad\;\; noise_1,\ldots,noise_{k-1})$
                    \STATE $noise_k = W\cdot\mathcal{N}(0,1)$
                    \STATE $Y_k = Y_k + \text{Synthesis}(u)$
                \ENDFOR
            \end{algorithmic}
        \end{minipage}
        }

        &

        \resizebox{.45\textwidth}{!}{
        \begin{minipage}{0.53\textwidth}
            \begin{algorithmic}[1]
                \REQUIRE Image $u$, level index $k$, steps $\mu$.
                \REQUIRE Images $Y_1,\ldots,Y_{k-1}$ (only for $k>1$).
                \REQUIRE Images $noise_1,\ldots,noise_{k-1}$ (only for $k>1$).
                \ENSURE Image $u$ (inplace)

                \IF{$k > 1$}
                    \FOR{$step = 1,\ldots,\mu$}
                        \STATE $d = BP^{\mu}_{k-1}($ \\
                            $\quad\quad \text{Downscale}(u), Y_1,\ldots,Y_{k-2},$ \\
                            $\quad\quad noise_1,\ldots,noise_{k-2}$ \\
                            $)$ \\
                        \STATE $u = u + \text{Upscale}(\left[Y_{k-1},\; d,\; noise_{k-1}\right])$
                    \ENDFOR
                \ENDIF
            \end{algorithmic}
            \end{minipage}
        }
    \end{tabular}
\end{algorithm}
\section{Generator Architecture}
\label{sec:generator}
\begin{figure}
    \centering
    \includegraphics[width=\linewidth]{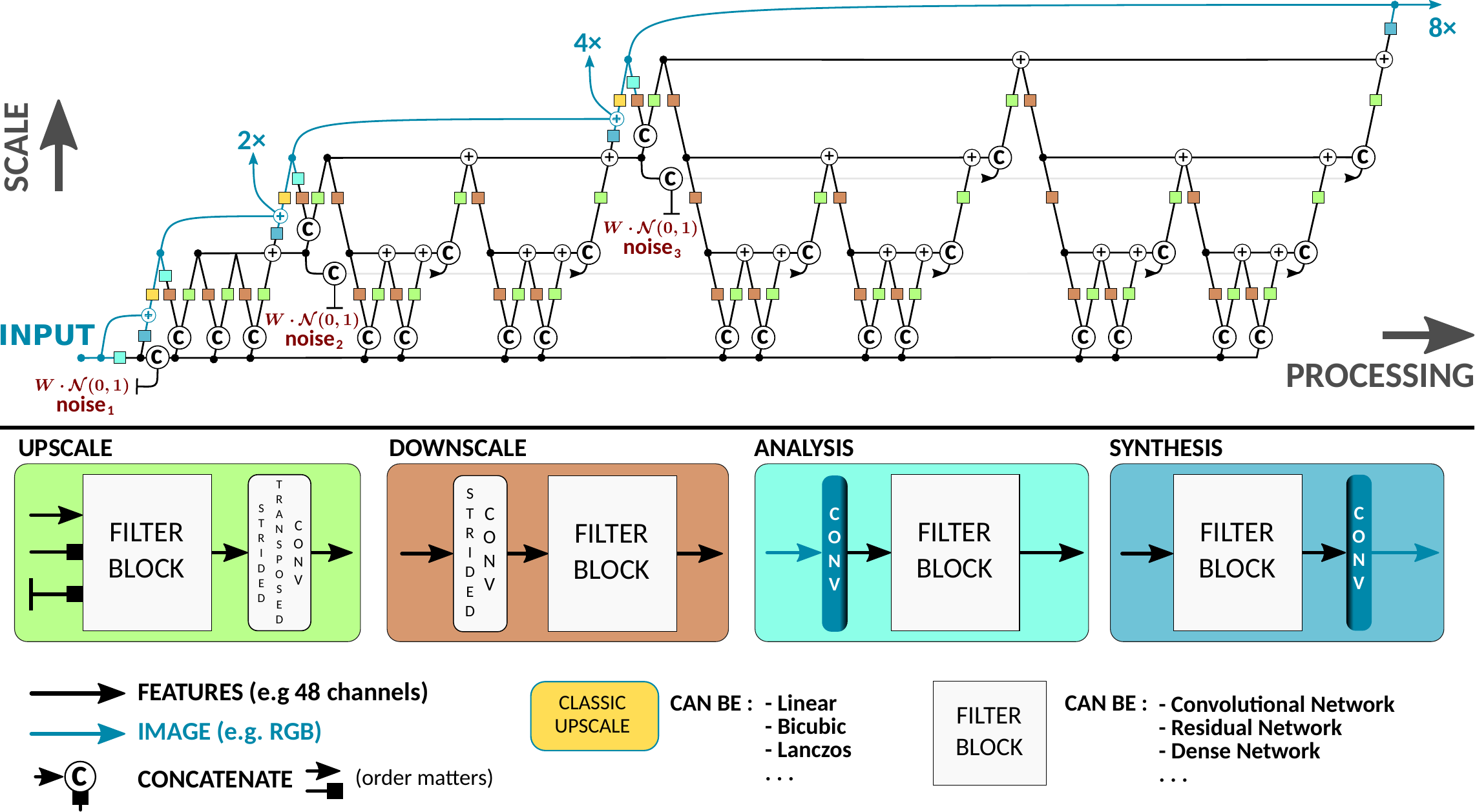}
    \caption{Generative Multi--Grid Back--Projection (G--MGBP) workflow, obtained from the recursion in Algorithm \ref{alg:gmgbp} with $\mu=2$ and $L=3$, to output $2\times$, $4\times$ and $8\times$ upscale images. One channel of $\mathcal{N}(0,1)$ noise enters each scale in feature space, and it is reused several times within each level. \label{fig:sys_workflow}}
\end{figure}

Our proposed architecture is shown in Figure \ref{fig:sys_workflow} and is based on the Multi--Grid Back--Projection (MGBP) algorithm from \cite{MGBP}, which improves a similar system used in NTIRE--SR Challenge 2018\cite{NTIRE2018_SR_Report}. This is a multi--scale super--resolution system that updates a progressive classic upscaler (like bicubic) with the output of a convolutional network system. At each level MGBP shares the parameters of all networks. The first upcale image at each level is obtained by a Laplacian pyramid approach\cite{MSLapSRN} and later improved by Iterative Back--Projections (IBP)\cite{Irani_1991a} computed in latent space (e.g. features within a network). Iterative Back--projections introduces a downscaler system to recover the low--resolution image from upscale images, and thus captures information from the acquisition model of the input image. By using back--projections in latent space, the downscaling model can be learned from training data and the iterations will enforce this model. For a multi--scale solution, MGBP uses a recursion based on multigrid algorithms\cite{UTrottenberg_2000a} so that, at each upscaling level, an image is updated recursively using all previous level outputs.

For the PIRM--SR Challenge 2018\cite{PIRM-SR} we extended MGBP to work as a generative system. For this purpose we added noise inputs that provide the \emph{innovation process} as explained in Section \ref{sec:strategy}. Previous work has shown the strong ability of convolutional networks to interpolate in feature space\cite{Radford2015UnsupervisedRL}. Inspired by this, we concatenate one channel of $\mathcal{N}(0,1)$ noise to the input features of the Upscaler module in Figure \ref{fig:sys_workflow}, and we use a parameter $W$ to control the amplitude of the noise. This parameter will later allow us to interpolate between distortion and perception optimizations (see Section \ref{ssec:pd}). In our experiments we use $48$ features, which increases to $49$ features with the noise input. The new recursive method is specified in Algorithm \ref{alg:gmgbp}.

The same noise channel is used during different IBP iterations at one scale ($\mu=2$ times in our experiments) and i.i.d. noise is used for different scales. Figure \ref{fig:sys_workflow} shows the unrolling recursion for $\mu=2$ number of back--projections.

\section{Discriminator Architecture}
\label{sec:discriminator}
\subsection{Variance Normalization and Shift Correlator}
\label{ssec:discr_normalization}
The task of the discriminator is to measure how realistic is an image. A straightforward approach is to input the color image to a convolutional network architecture. Then, we hope that the discriminator learns from adversarial training using real and fake image examples. In practice, we find that this approach works well to identify which areas of upscale images need more textures but the artificial details look noisy and have limited structure.

So what makes an image look natural? Extensive research has been carried to address this question. Here, we follow the seminal work of Ruderman\cite{Ruderman1994} who found regular statistical properties in natural images that are modified by distortions. In particular, Ruderman observed that applying the so--called \emph{variance normalization} operation:
\begin{equation}
    \hat{I}_{i,j} = \frac{I_{i,j}-\mu_{i,j}(I)}{\sigma_{i,j}(I)+1} \;,
\end{equation}
has a decorrelating effect on natural images. Here, $I_{i,j}$ is the luminance channel of an image with values in $[0, 255]$ at pixel $(i,j)$, $\mu(I)$ is the local mean of $I$ (e.g. output of a Gaussian filter), and $\sigma(I)^2=\mu(I^2)-\mu^2(I)$ is the local variance of $I$. Ruderman also observed that these normalized values strongly tend towards a Gaussian characteristic for natural images. These findings are used in the NIQE perceptual quality metric considered for the PIRM--SR Challenge 2018\cite{NIQE_2013}. NIQE also models the statistical relationships between neighboring pixels by considering horizontal and vertical neighbor products: $\hat{I}_{i,j}\hat{I}_{i,j+1}$, $\hat{I}_{i,j}\hat{I}_{i+1,j}$, $\hat{I}_{i,j}\hat{I}_{i,j-1}$ and $\hat{I}_{i,j}\hat{I}_{i-1,j}$. Previously, the BRISQUE non--reference metric also used diagonal products\cite{BRISQUE_2012}.

\begin{figure}[t]
    \centering
    \includegraphics[width=0.8\linewidth]{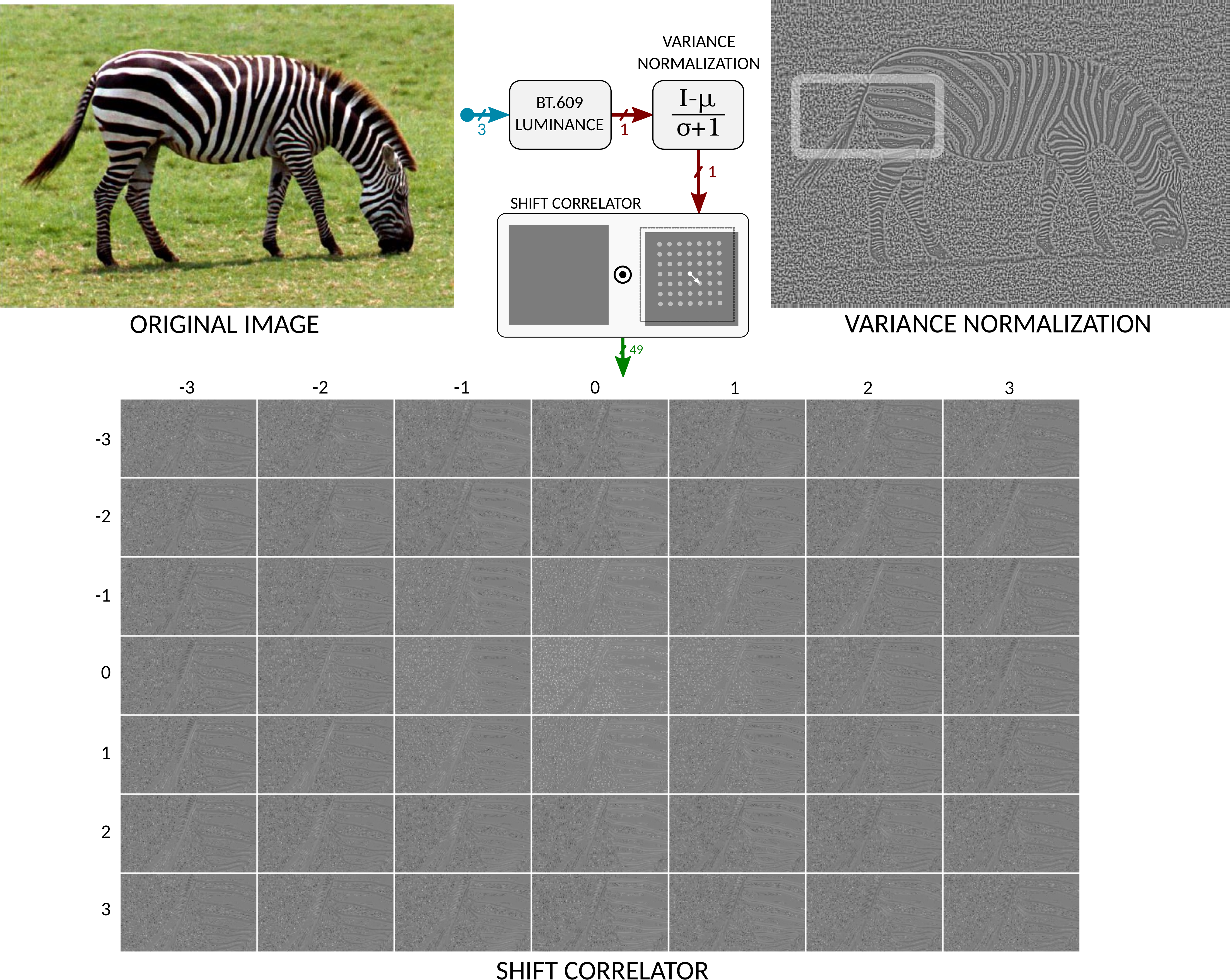}
    \caption{We propose a Variance Normalization and Shift Correlator (VN+SC) layer that transforms the inputs of the discriminator into $49$ channels that, according to research on the statistics of natural images, capture the essential information to discriminate between natural and unnatural images. \label{fig:vnsc}}
\end{figure}

\begin{figure}
    \centering
    \includegraphics[width=\linewidth]{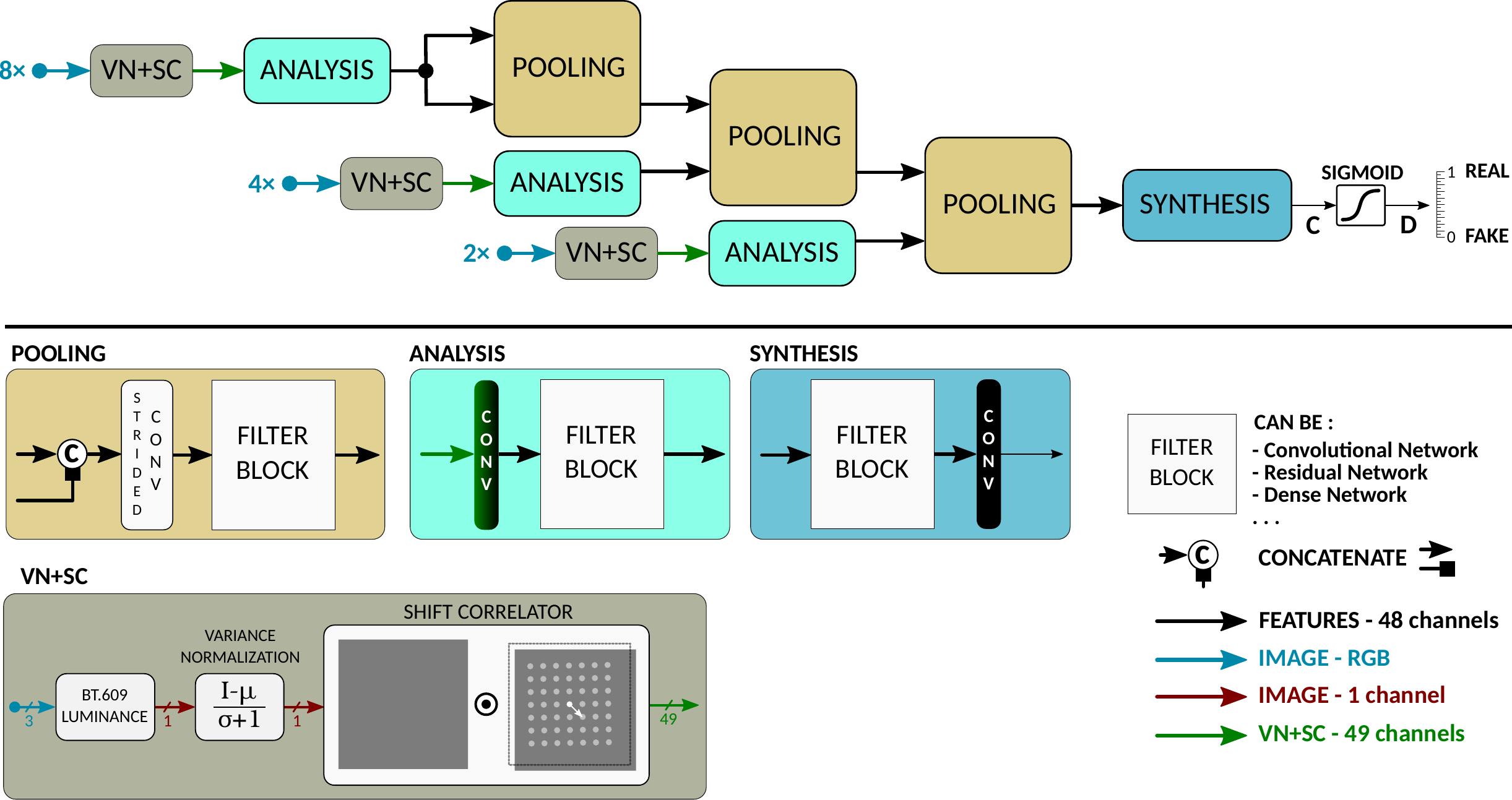}
    \caption{Multi--level recursive discriminator used for adversarial training. The diagram shows $D^3$, the system unfold for $3$ levels to simultaneously evaluate $2\times$, $4\times$ and $8\times$ upscale outputs. Each module shares parameters in different scales. \label{fig:sys_discriminator}}
\end{figure}

Inspired by previous research we define the Variance Normalization and Shift Correlator (VN+SC) layer as follows:
\begin{equation}
    V^{7(p+3)+q+3}_{i,j}(I) = \hat{I}_{i,j}\cdot\hat{I}_{i+p,j+q} \;, \quad\quad p=-3,\ldots,3\quad,q=-3,\ldots,3\;.
\end{equation}
Here, we transform a color image into a set of neighbor products (shift correlator) $V^k_{i,j}$ with $k=0,\ldots,48$, using the variance normalized image $\hat{I}$. The number of neighbor products can be any number, and we set it to $7\times7$ in our experiments to get a number similar to the $48$ features used in our discriminator architecture. Figure \ref{fig:vnsc} shows the visual effect of the the VN+SC operation. We use a VN+SC layer for each input of our discriminator, as shown in Figure \ref{fig:sys_discriminator}.


\subsection{Multi--Scale and Recursive Architecture}
\label{ssec:discr_multiscale_recursive}
The G--MGBP upscaler designed in Section \ref{sec:generator} is multi--scale and recursive. We can then take advantage of the multi--scale distortion optimization training strategy proposed in \cite{MSLapSRN}. This strategy is difficult for adversarial training because the outputs at different levels contain different artifacts and might need an ensemble of discriminators. We simplify this problem by using a multi--scale and recursive architecture as shown in Figure \ref{fig:sys_discriminator}. The system takes several upscaled images using different factors ($2\times$, $4\times$ and $8\times$ in our experiments) and, based on all of them, it outputs one score to decide if the images are real or fake. The parameters of each block (pooling, analysis and synthesis) are shared at each level. Thus, the system keeps the same number of parameters, either in a small configuration with $L=1$ level to evaluate a single $2\times$ upscale output, or in a large configuration with $L=3$ levels to simultaneously evaluate $2\times$, $4\times$ and $8\times$ upscale outputs. Adversarial training with this discriminator resembles a Progressive GAN\cite{karras2018progressive} because it can adjust parameters to first solve the simpler problem of $2\times$ upscaling, and then follow with larger factors. But, at the same time, it is significantly different because a Progressive GAN system is neither multi--scale nor recursive.

\section{Adversarial Training Strategy}
\label{sec:training}
We follow the design of multi--scale loss from MSLapSR\cite{MSLapSRN} with $3$ scales: $2\times$, $4\times$ and $8\times$. For each scale $L\in\{1,2,3\}$ we take $X^L$, as patches from the HR dataset images. High--resolution references $X^k$ with $k=1,\ldots,L-1$ are obtained by downscaling the dataset HR images with factor $L-k$. This is:
\begin{eqnarray}
    X^L & = & \text{HR image from dataset} \;, \quad L=1\,,2\,,3 \;, \\
    X^k & = & S^{L-k}_{down}(X^L) \;, \quad k=1,\ldots,L-1 \;.
\end{eqnarray}
We denote $Y_{W=0}$ and $Y_{W=1}$ the outputs of our generator architecture using noise amplitudes $W=0$ and $W=1$, respectively. Then, we combine the multi--scale loss from \cite{MSLapSRN} and the perceptual loss from \cite{mechrez2018Learning} with different noise inputs. Our total loss is given by:
\begin{eqnarray}
    \label{eq:total_loss}
    \mathcal{L}(Y, X; \theta) = \sum_{L=1,2,3} \Big\{ & & 0.001 \cdot {\color{mygreen}\mathcal{L}^{GAN-G}_L(Y_{W=1})} + 0.1 \cdot {\color{mygreen}\mathcal{L}^{context}_L(Y_{W=1}, X)} + \notag\\
                                                      & & 10 \cdot {\color{blue}\mathcal{L}^{rec}_L(Y_{W=0}, X)} + 10 \cdot {\color{red}\mathcal{L}^{cycle}_L(Y_{W=0}, Y_{W=1}, X)}\Big\} \;.
\end{eqnarray}
\begin{figure}[t]
    \centering
    \includegraphics[width=0.8\linewidth]{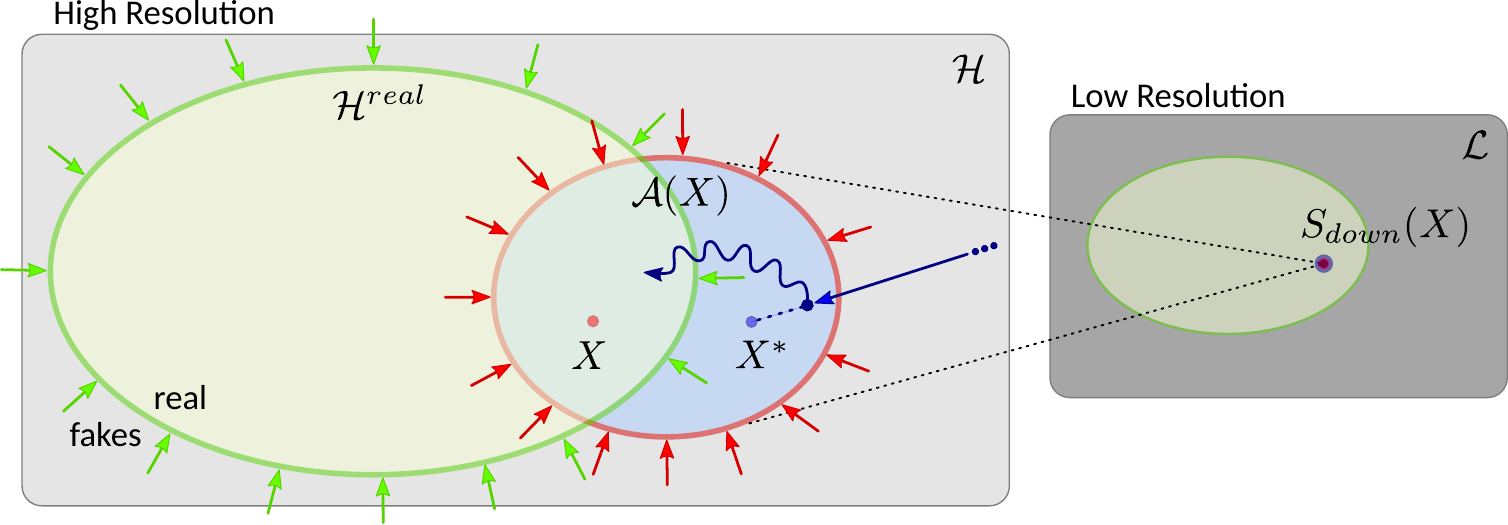}
    \caption{Our loss function tries to: look real by moving into $\mathcal{H}^{real}$ (GAN and CX loss), enforce a downscaling model by moving into $\mathcal{A}(X)$ (cycle loss), and be reachable by latent space interpolation from the optimal distortion solution $X^*$ (reconstruction loss). \label{fig:strategy_losses}}
\end{figure}
Here, colors represent the target of each loss term according to Figure \ref{fig:strategy_losses}. First,
\begin{eqnarray}
    {\color{mygreen}\mathcal{L}^{GAN-G}_L(Y_{W=1})} & = & \mathbb{E}\left[ \log(D^L(Y^k_{W=1} | k=1,\ldots,L) \right] \;, \\
    {\color{mygreen}\mathcal{L}^{GAN-D}_L(Y_{W=1})} & = & \mathbb{E}\left[ \log(D^L(X^k | k=1,\ldots,L)) \right] + \notag \\
                                                    &   & \mathbb{E}\left[ \log(1-D^L(Y^k_{W=1} | k=1,\ldots,L)) \right]
\end{eqnarray}
follows a standard adversarial loss\cite{NIPS2014_5423}, where $D^L$ is our $L$--level recursive discriminator evaluating $L$ output images, as shown in Figure \ref{fig:sys_discriminator}. Then,
\begin{equation}
    {\color{mygreen}\mathcal{L}^{context}_L(Y_{W=1}, X)} = - \mathbb{E}\left[ \sum_{k=1}^L \log\left(CX(\Phi(Y^k_{W=1}), \Phi(X^k)\right) \right]
\end{equation}
uses the \emph{contextual similarity} $CX$ as defined in \cite{mechrez2018contextual} and $\Phi$ are features from \emph{conv3--4} of VGG--19 network as suggested in \cite{mechrez2018Learning}. The contextual loss is designed to give higher importance to the perceptual quality\cite{mechrez2018contextual}. Next,
\begin{equation}
    {\color{blue}\mathcal{L}^{rec}_L(Y_{W=0}, X)} = \mathbb{E}\left[ \sum_{k=1}^L ||Y^k_{W=0} - X^k||_1 \right]
\end{equation}
is a standard $L1$ distortion loss, equivalent to the multi--scale loss in \cite{MSLapSRN}. We note that here the noise input is set to zero, which prevents this term to interfere with the generation of details as it does not see randomness in the outputs. Finally, the \emph{cycle} regularization loss enforces the downscaling model by moving the outputs back to low--resolution, analogous to the cycle--loss in CycleGAN\cite{CycleGAN2017}. This is,
\begin{eqnarray}
    {\color{red}\mathcal{L}^{cycle}_L(Y_{W=0}, Y_{W=1}, X)} & = & \mathbb{E}\left[ \sum_{k=1}^L \sum_{f=1}^k ||S^f_{down}(Y^k_{W=0})-S^f_{down}(X^k)||_1 \right] + \\
                                                            &   & \mathbb{E}\left[ \sum_{k=1}^L \sum_{f=1}^k ||S^f_{down}(Y^k_{W=1})-S^f_{down}(X^k)||_1 \right] \;,
\end{eqnarray}
where we use the $L1$ distance between downscaled outputs and low--resolution inputs. The first term, with noise amplitude zero, forces $Y_{W=0}$ to stay in $\mathcal{A}(X)$ as it approaches the image $X^*$. The second term, with unit noise, forces $Y_{W=1}$ to stay in $\mathcal{A}(X)$ as it approaches the set $\mathcal{H}^{real}$.

\section{Experiments}
\label{sec:experiments}

\subsection{Configuration}
\label{ssec:configuration}
For training and validation data we resized images to the average mega--pixels of PIRM--SR dataset ($0.29$ Mpx), taking all images from: DIV2K\cite{Agustsson_2017_CVPR_Workshops}, FLICKR--2K, CLIC (professional sets), and PIRM--SR self--validation\cite{PIRM-SR}. We selected $4,271$ images for training and $14$ images for validation during training.

We used one single configuration to test our system. We configure the \emph{Analysis}, \emph{Synthesis}, \emph{Upscale}, \emph{Downscale} and \emph{Pooling} modules in Figure \ref{fig:sys_workflow} and \ref{fig:sys_discriminator} using $4$--layer dense networks\cite{huang2017densely}  as filter--blocks. We use $48$ features and growth rate $16$ within dense networks. For classic upscaler we started with Bicubic and we set the upscaling filters as parameters to learn as proposed in \cite{MSLapSRN}.

We trained our system with Adam optimizer and a learning rate initialized as $10^{-3}$ and square root decay, both for generator and discriminator systems. We use $128\times128$ patches with batch size $16$. We pre--trained the network with $W=0$ and only L1 loss, and used as initial setting for our overall loss \eqref{eq:total_loss}.

\subsection{Moving on the Perception--Distortion plane}
\label{ssec:pd}
\begin{figure}[t]
    \centering
    \includegraphics[width=\linewidth]{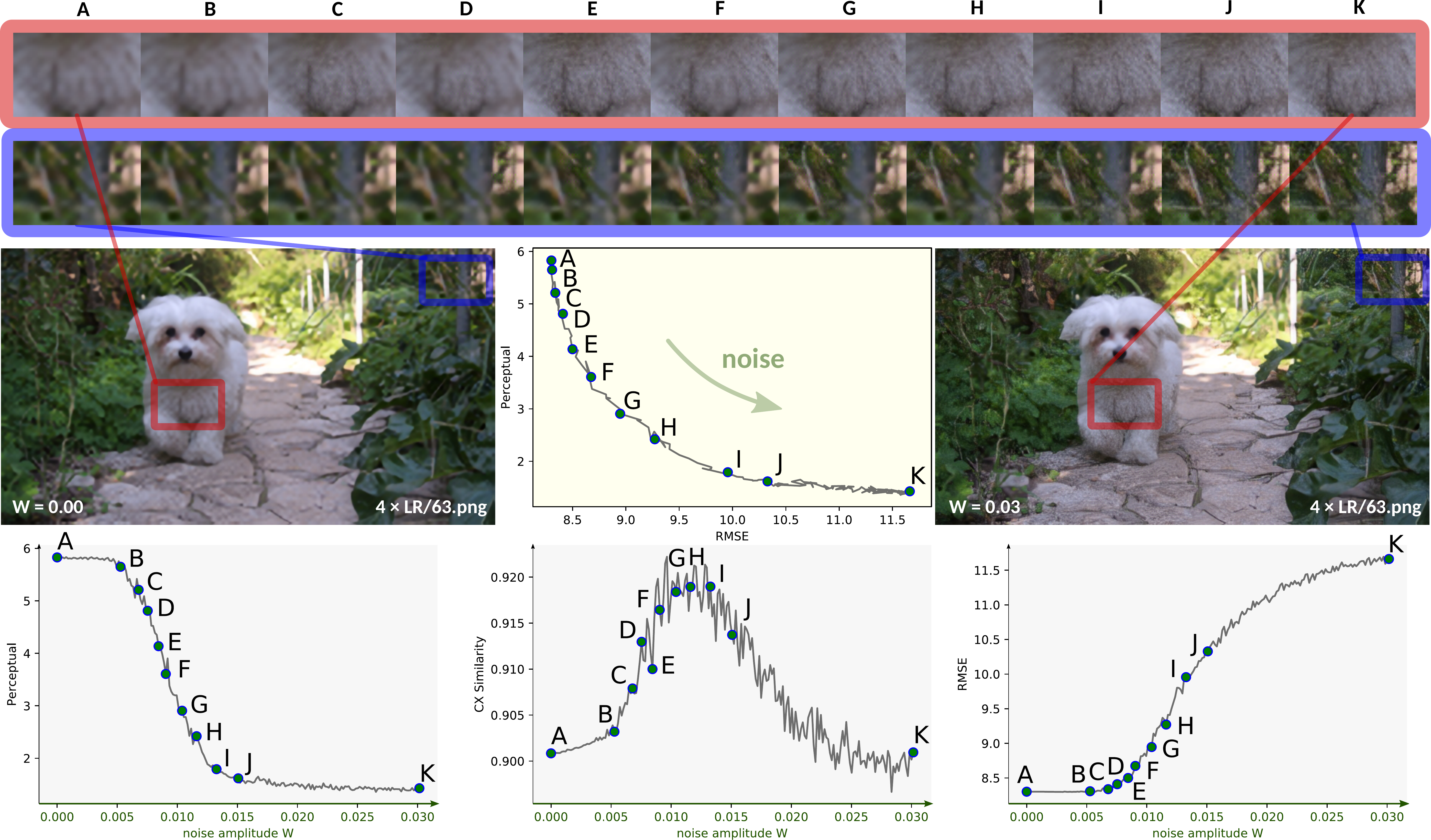}
    \caption{Our SR architecture uses noise inputs that can be use to move from distortion to perception optimization, without retraining the system. The plot shows how perceptual quality improves as we increase the noise amplitude in our R3 model. Output images show how artificial details appear in different areas of the image as noise amplifies. \label{fig:interpolation}}
\end{figure}
An essential part of our generative SR architecture is the noise inputs. The training strategy introduced in Section \ref{sec:training} teaches the system to optimize distortion when the noise is set to zero, and maximize perceptual quality when the noise is enabled. Thus, noise provides the randomness needed for natural images and represents the \emph{innovation jump} according to Figure \ref{fig:strategy_sets}.

After training, we are free to control the noise inputs. In particular, we can move the noise amplitude smoothly between $W=0$ and $W=1$ to inspect the path to jump from distortion to perception optimization. Figure \ref{fig:interpolation} shows an example of this transition. Here, it is important to note that our training strategy does not optimize the trajectory in the perception--distortion plane, but only the corner cases of best distortion ($W=0$) and best perception ($W=1$). The corner cases are clearly verified in Figure \ref{fig:interpolation}. At this point, it is unkown which trajectory will the the network take to move from one case to the other.

It is interesting to see in Figure \ref{fig:interpolation} that the transition from best perception to best distortion happens within a narrow margin of $\Delta W = 0.02$ amplitude values and much closer to $W=0$ than $W=1$ (around $W\sim0.01$). Similar transitions were observed in other images of the PIRM dataset, for both test and validation.

We also observe that the parametric curve in the perception--distortion plane looks like a monotonically non--increasing and convex function, similar to the optimal solution studied in \cite{Blau_2018_CVPR}. But, it is important to emphasize that the curve in Figure \ref{fig:interpolation} is not optimal as we are not enforcing optimality and, as a matter of fact, for the PIRM--SR Challenge we ended up using different training results for R1, R2 and R3, each one performing better than the others in its own region.

Regarding image quality metrics, we see with no surprise that the \emph{Perceptual} index proposed for the PIRM--SR Challenge\cite{PIRM-SR} improves as noise increases, while the distortion measured by RMSE increases. We observed very similar results for the perceptual metrics NIQE and Ma, as well as the L1 distortion metric. More interesting is the transition observed in the \emph{contextual similarity} index. First, it behaves as a perceptual score with the CX similarity improving consistently as noise increases. Then, when the \emph{Perceptual} score seems to stall, but RMSE keeps increasing, the CX similarity changes to a distortion metric pattern, reducing as noise increases. This is consistent with the design target of \emph{CX similarity} to focus more on perceptual quality while maintaining a reasonable level of distortion\cite{mechrez2018contextual}.

\subsection{Ablation Tests}
\label{ssec:ablation}
\begin{figure}
    \centering
    \includegraphics[width=\linewidth]{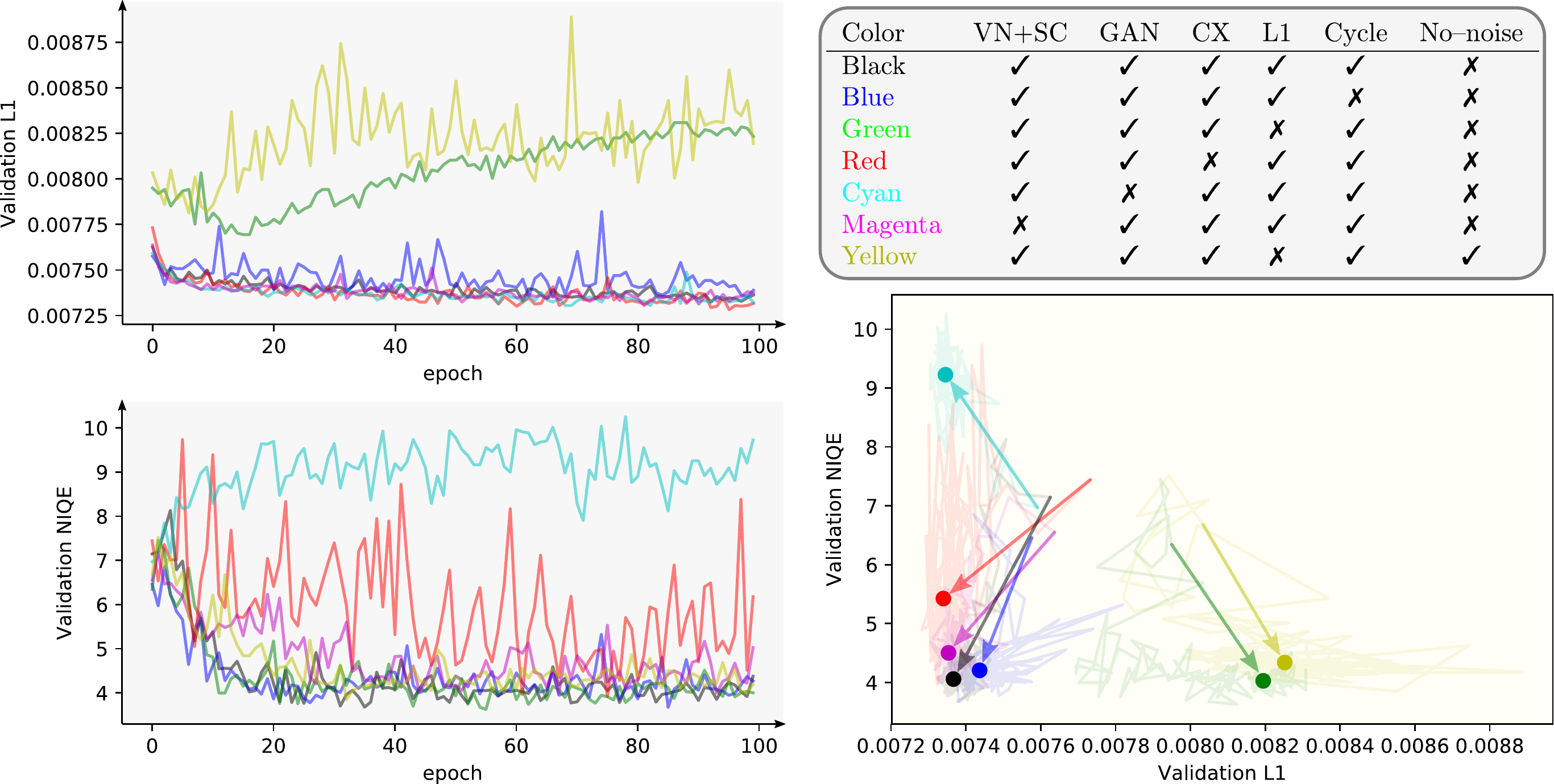}
    \caption{Ablation tests show the validation scores when training our network for $100$ epochs. We consider removal of the loss terms: GAN, CX, L1 and Cycle in \eqref{eq:total_loss}, as well as VN+SC layers in the discriminator, and training the system without noise inputs.}
    \label{fig:ablation}
\end{figure}
Our overal loss combines terms focused on different targets (e.g. low distortion, perceptual quality). In Section \ref{sec:training} we explained the purpose of each term using the diagram in Figure \ref{fig:strategy_sets}. It remains to verify this design and to quantify the relevance of each term. We also want to quantify the contribution of our novel VN+SC layer. For this purpose we trained our network architecture for $100$ epochs according to the configuration in section \ref{ssec:configuration}. In Figure \ref{fig:ablation} we show our measurements of L1 (distortion) and NIQE (perceptual) in a small validation set of $14$ images after each epoch. We display the evolution through the number of epochs as well as the trajectories on the perception--distortion plane.

Overall, we see that our strategy adding all the losses (in black color) gives the best perception--distortion balance. In the extremes we see that removing the L1 and GAN losses have catastrophic effects on distortion and perception, respectively. Still, these cases do not diverge to infinity because of other loss terms. Next, it is clear that the contextual loss helps improving the perceptual quality, and regarding distortion the amount of improvement is not conclusive. Then, the addition of the cycle loss shows a clear improvement over distortion, with unconclusive improvements on perceptual quality. And finally, we observe that the addition of the VN+SC layer in the discriminator clearly improves perceptual quality, although not as much as CX and GAN losses.

Figure \ref{fig:ablation} also shows a test in which we avoid the use of noise imputs by setting $W=0$ in all losses. In this case we remove the L1 loss that would otherwise interfere with the GAN loss, causing a catastrophic effect. In this case distortion is controlled by the cycle loss, equivalent to how it is done in \cite{mechrez2018Learning}. In this configuration the network performs slightly worse in perceptual quality and clearly worse on distortion, similar to only removing the L1 loss. In this case, we believe that the network uses the randomness in the input as innovation process, which cannot be controlled and limits the diversity of the generator.

\subsection{Challenge Results}
\label{ssec:results}
Table \ref{tab:results} shows our best average scores in the PIRM--SR Challenge 2018\cite{PIRM-SR} for Region 1 ($RMSE\leqslant 11.5$), Region 2 ($11.5<RMSE\leqslant 12.5$) and Region 3 ($12.5<RMSE\leqslant 16$), compared to baseline methods: EDSR\cite{Lim_2017_CVPR_Workshops}, CX\cite{mechrez2018contextual} and EnhanceNet\cite{Sajjadi_2017_ICCV}. We achieved better perceptual scores compared to all baselines.
\begin{table}[t]
    \caption{Quantitative comparison between our solutions for R1, R2 and R3 and baseline methods in the test set. Best numbers in each row are shown in bold.}
    \label{tab:results}
    \centering
    \setlength{\tabcolsep}{4pt}
    \begin{tabular}{lcrrrrrr}
        \hline
                   &                                             & EDSR                  &                                        \multicolumn{3}{c}{Our} &      CX &    ENet \\
                   & Unit/Metric                                 &                       &                 R1 &                 R2 &                   R3 &         &         \\ \hline
        Parameters &  [k]                                        & $43,100$              & $\boldsymbol{281}$ & $\boldsymbol{281}$ &   $\boldsymbol{281}$ &   $853$ &   $853$ \\
        Perceptual &  $\tfrac{1}{2}((10-\text{Ma})+\text{NIQE})$ &  $4.904$              &            $3.817$ &            $2.484$ & $\boldsymbol{2.019}$ & $2.113$ & $2.723$ \\
        Distortion &  $RMSE$                                     &  $\boldsymbol{10.73}$ &            $11.50$ &            $12.50$ &              $14.24$ & $15.07$ & $15.92$ \\ \hline
    \end{tabular}
\end{table}
\begin{figure}
    \centering
    \includegraphics[width=0.8\linewidth]{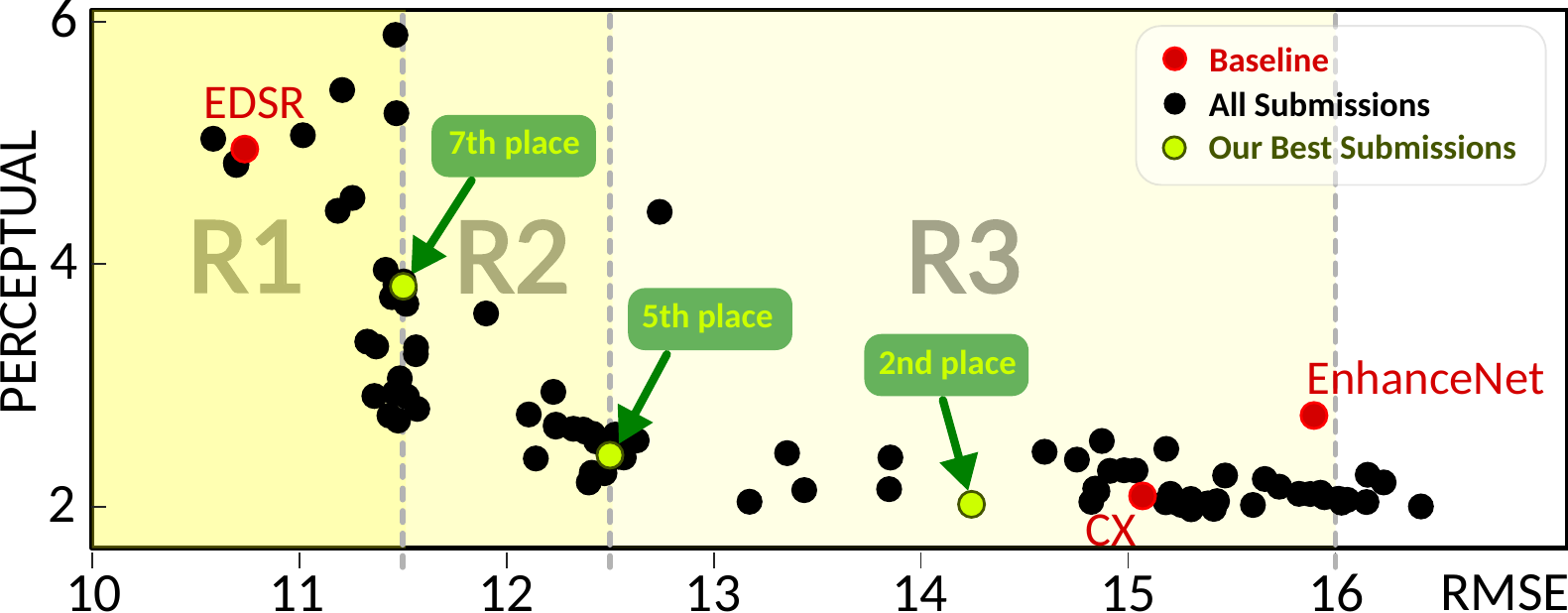}
    \caption{Perception--distortion plane with average scores in the test set showing all submissions, from all teams in PIRM--SR 2018\cite{PIRM-SR}. Our best scores are shown in green color together with the final ranking in PIRM--SR Challenge 2018. \label{fig:pirm2018_final_results}}
\end{figure}

Beyond the target of the competition, we also observe that we use significantly less parameters. This shows the advantage of the recursive structure of our system, which successfully works across multiple scales to achieve the target. Our system can upscale one image of the self--validation set in $0.2s$ in average.

Compared to other submissions, we observe in Figure \ref{fig:pirm2018_final_results} that our system performs better in Region 3. Here, we achieve the $2^{nd}$ place within very small differences in perceptual scores but with significantly lower distortion. This shows the advantage of our training strategy to optimize the perception--distortion trade--off. In Regions 1 and 2 we were one among only two teams that reached the exact distortion limit ($11.5$ in Region 1 and $12.5$ in Region 2). We were able to achieve this by controlling the noise amplitude, without retraining the system. Our ranking lowers as the distortion target gets more difficult. We believe that this is caused by the small size of our system that becomes more important for low distortion targets, since we use only $281k$ parameters compared to $43M$ of the EDSR baseline in Region 1.

\begin{figure}
    \centering
    \includegraphics[width=0.92\linewidth]{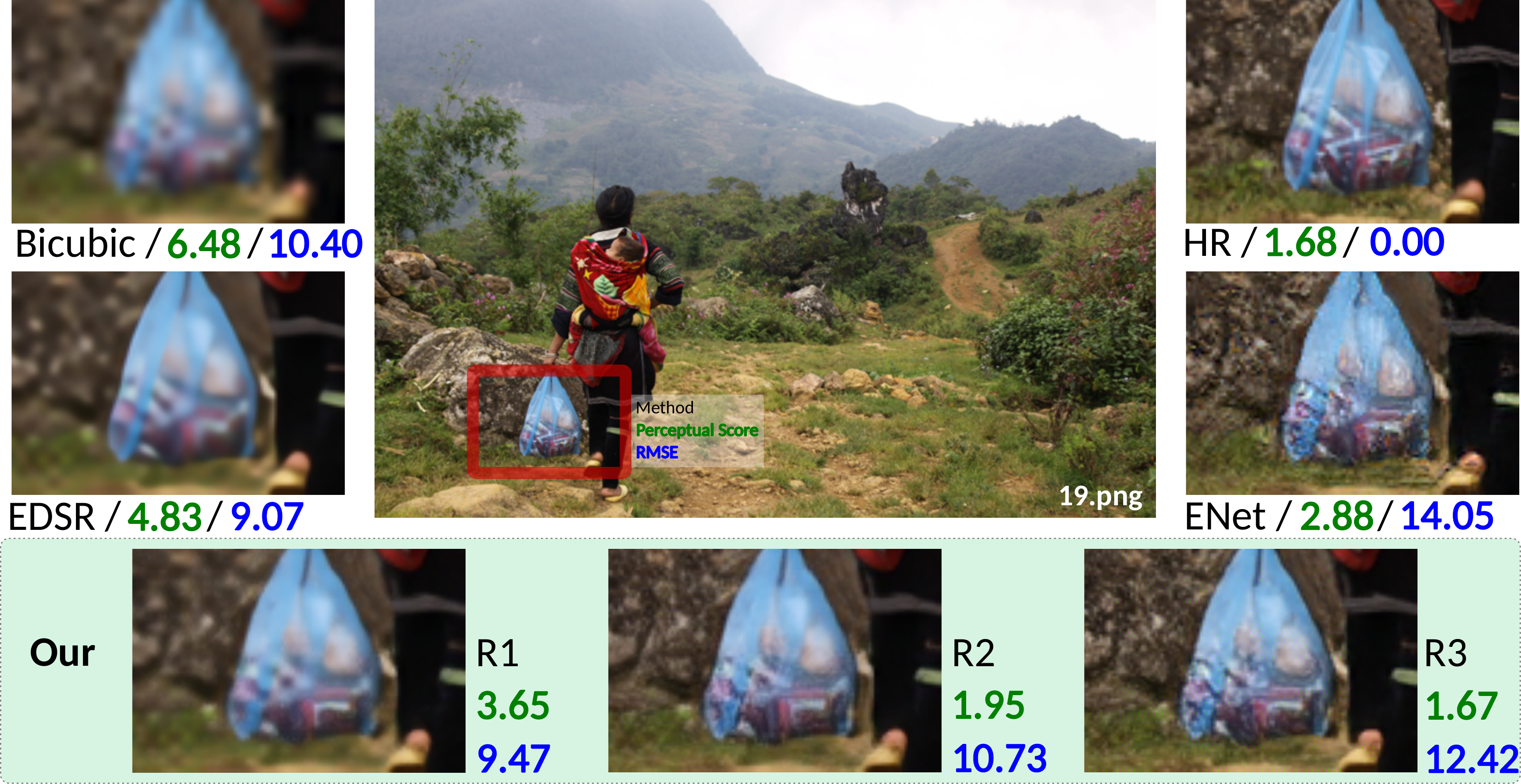} \medskip
    \hrule \medskip
    \includegraphics[width=0.92\linewidth]{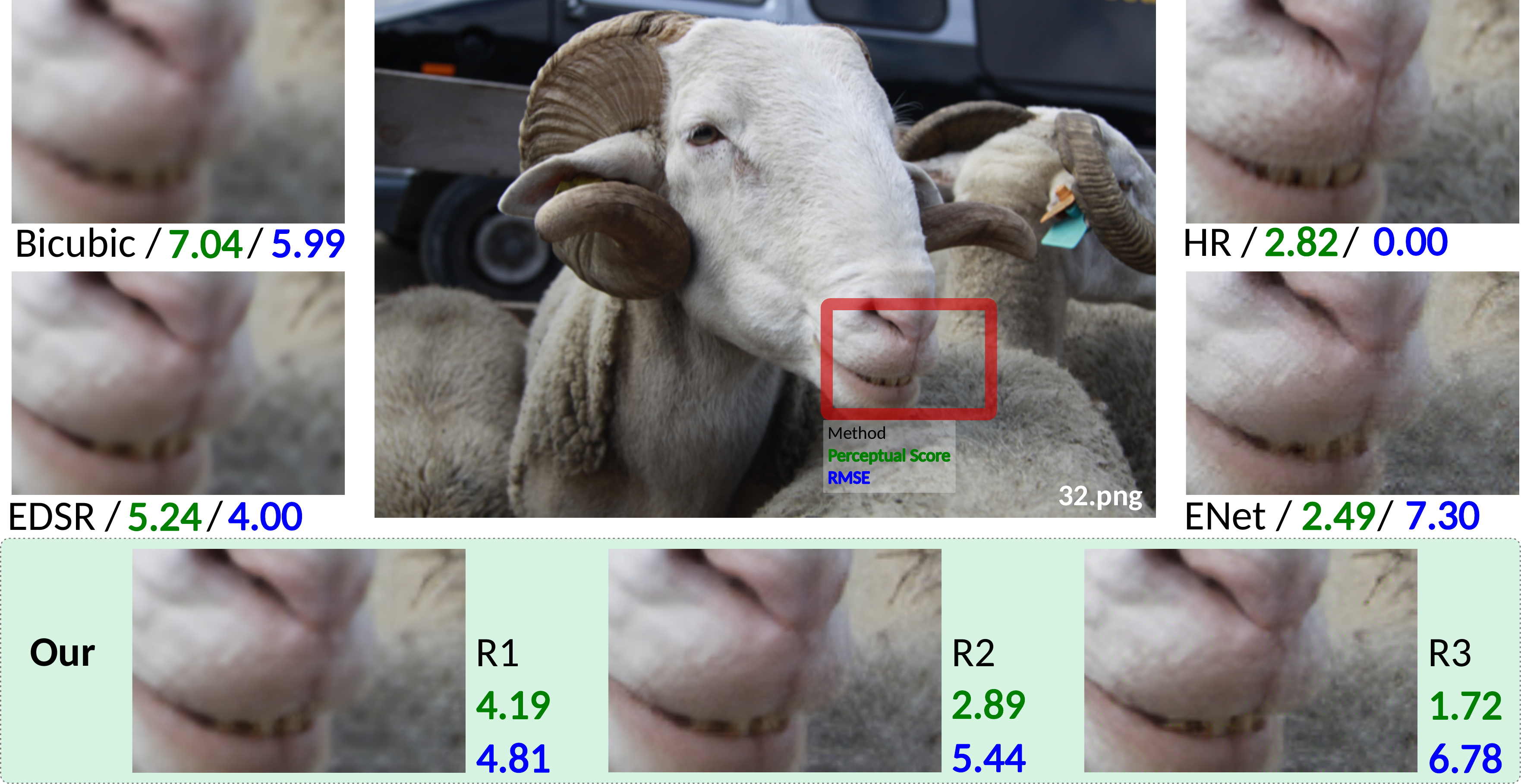} \medskip
    \hrule \medskip
    \includegraphics[width=0.92\linewidth]{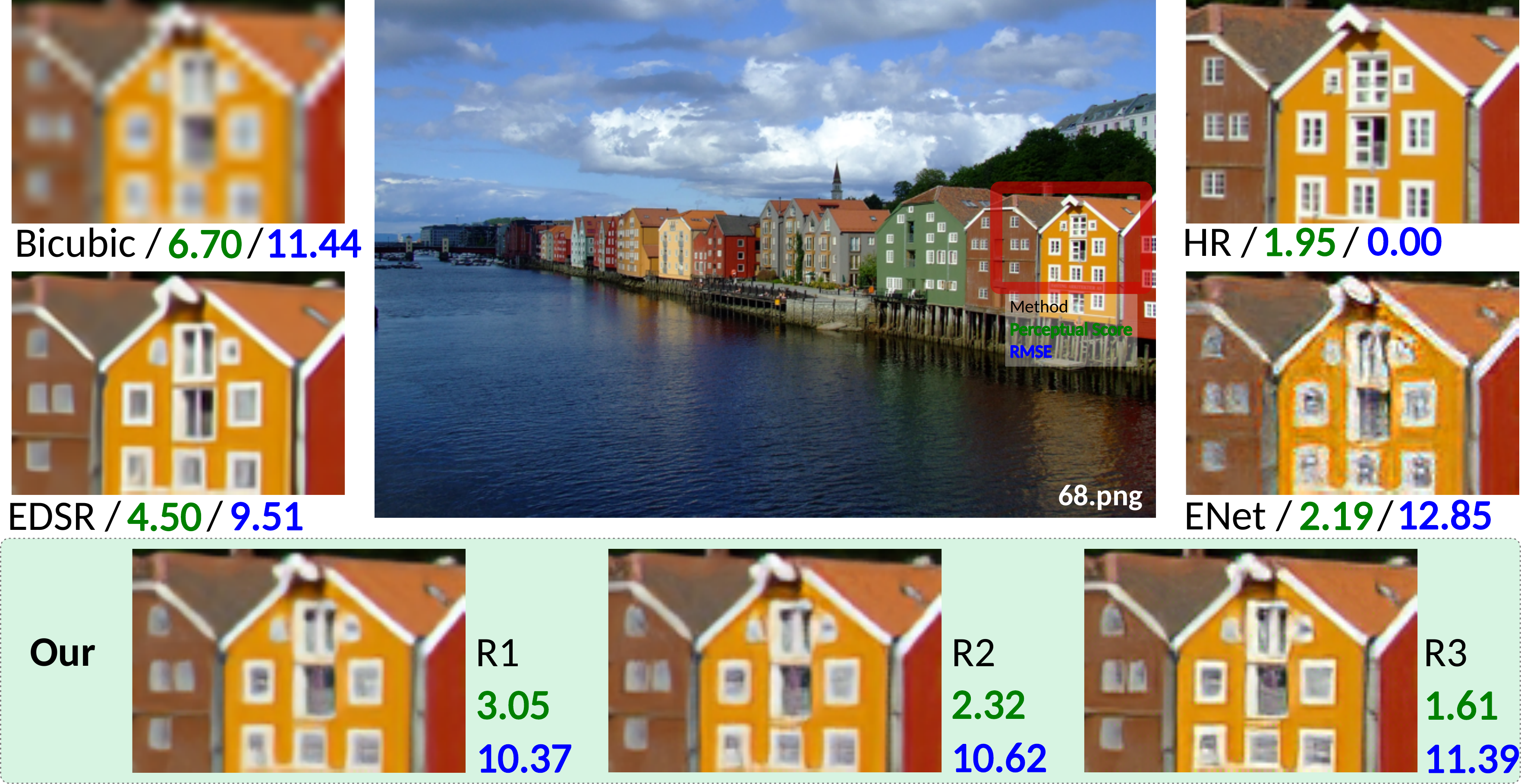}
    \caption{Image comparisons of $4\times$ upscaling between our solutions in R1, R2 and R3 (see Figure \ref{fig:pirm2018_final_results}) and baseline methods in our validation set. Perceptual and distortion scores of whole images are shown in green and blue colors, respectively. \label{fig:comparison}}
\end{figure}

Finally, Figure \ref{fig:teaser} and \ref{fig:comparison} show comparisons of our results with the baselines, using images from our validation set. We observe that in Region 3 we achieve better perceptual scores even compared to the original HR images. While we subjectively confirm this in some patches, we do not make the same conclusion after observing the whole images. Somehow, we believe that our design for adversarial training and validation strategy managed to overfit the perceptual scores. Nevertheless, we observe clear advantages to the baselines, showing better structure in textures and more consistent geometry in edges and shapes.

%
%
%
\bibliographystyle{splncs04}
\bibliography{paper}
\end{document}